\documentclass[preprint,aps,showpacs,showkeys,superscriptaddress]{revtex4}
\usepackage[latin9]{inputenc}
\usepackage{color}
\usepackage{textcomp}
\usepackage{amsmath}
\usepackage{amssymb}
\usepackage{graphicx}

\makeatletter
\@ifundefined{textcolor}{}
{%
 \definecolor{BLACK}{gray}{0}
 \definecolor{WHITE}{gray}{1}
 \definecolor{RED}{rgb}{1,0,0}
 \definecolor{GREEN}{rgb}{0,1,0}
 \definecolor{BLUE}{rgb}{0,0,1}
 \definecolor{CYAN}{cmyk}{1,0,0,0}
 \definecolor{MAGENTA}{cmyk}{0,1,0,0}
 \definecolor{YELLOW}{cmyk}{0,0,1,0}
}

\usepackage{bm}
\usepackage{indentfirst}
\usepackage{float}
\usepackage{caption}
\usepackage{psfrag}
\usepackage{epstopdf}
\usepackage{diagbox}
\usepackage{orcidlink}

\makeatother
\begin{document}
\title{Probing Short-Distance Modifications of Gravity via Spin-Independent and
Spin-Dependent Effects in Muonic Atoms}
\author{J. E. J. Matias}
\affiliation{Department of Physics, Federal University of Paraiba, Jo\~{a}o Pessoa 58051-900,
PB, Brazil}
\author{A. S. Lemos}
\email{adiel@ufersa.edu.br}
\affiliation{Departamento de Ci\^{e}ncias Exatas e Tecnologia da Informa\c{c}\~{a}o, Universidade
Federal Rural do Semi-\'{A}rido, 59515-000 Angicos, Rio Grande do Norte, Brazil}
\author{F. Dahia}
\email{fdahia@fisica.ufpb.br}
\affiliation{Department of Physics, Federal University of Paraiba, Jo\~{a}o Pessoa 58051-900,
PB, Brazil}


\begin{abstract}
High-precision spectroscopy of muonic atoms provides a powerful probe for new
short-range interactions predicted by theories beyond the Standard Model (SM).
In this work, we derive new constraints on both spin-independent and
spin-dependent non-Newtonian gravity by leveraging the outstanding sensitivity
of these systems. For spin-independent Yukawa-type forces, we analyze two
complementary approaches: the $2S-2P$ Lamb shift in the muonic helium-4 ion
and the deuteron-proton squared charge radii difference obtained from the
muonic hydrogen-deuterium isotope shift. The found constraints have reached a
competitive level at sub-picometer scales, with the isotope shift method
yielding the most stringent bounds for interaction ranges $\lambda
\lesssim10^{-13}\text{ m}$. For spin-dependent effects, we analyze the
influence of the gravitational spin-orbit coupling on the $2P_{3/2}-2P_{1/2}$
fine-structure splitting in muonic helium, establishing new limits on
Post-Newtonian parameters. These bounds are shown to be more restrictive than
those from other leading experimental techniques for ranges $\lambda
\lesssim10^{-10}\text{ m}$. Our findings highlight the widespread usefulness
of muonic atoms in exploring new fundamental physics at short-distance scales.

\end{abstract}
\maketitle

\section{Introduction}

Modern physics is built upon two pillars, the Standard Model (SM) of particle
physics and General Relativity. Despite their significant achievements, a
complete unification remains elusive, and foundational issues such as the
hierarchy problem suggest the existence of new physics beyond our current
understanding. One compelling theoretical proposal involves the existence of
large extra spatial dimensions, which, from lowering the fundamental Planck
scale, could solve the hierarchy problem \citep{Hamed1998,add2,rs1,rs2}. A
generic prediction of these theories is the emergence of new, short-range
interactions that would manifest as deviations from Newtonian gravity at the
microscopic scale \citep{kehagias,colliders}.

The experimental search for such deviations from standard gravity covers a
wide range of energy and distance scales. On astrophysical and cosmological
scales, observations of phenomena such as light deflection and the Shapiro
delay place stringent bounds on Post-Newtonian parameters
\citep{Will:2014kxa}. At terrestrial and microscopic scales, the search for
new interactions is often framed in terms of a Yukawa-type correction to the
Newtonian potential, characterized by strength $\alpha$ and range $\lambda$. A
dedicated experimental program, including high-precision tests with torsion
balances \citep{hoyle01,hoyle04,hoyle07,adelbergREV,Lee:2020zjt,radion} and
measurements of the Casimir effect
\citep{Klimchitskaya:2024ttt,Klimchitskaya:2023niz,murata}, has relentlessly
pushed the limits on such new forces. Complementary to these direct
gravitational probes, atomic and molecular spectroscopy offers an independent
and robust method for searching for new physics
\citep{molecule,safranova,lemos1,lemos2,lemos3,lemos4,lemos5,Dahia2024,atomicspec1,atomicspec2,atomicspec3}.
By measuring energy-level shifts with remarkable accuracy, atomic systems can
reveal the subtle influence of new particles or interactions. At the same
time, experiments probing fundamental symmetries and spin-gravity couplings,
such as the MTV-G experiment \citep{Tanaka:2014jfa,Tanaka:2013ika}, can
provide another avenue for discovering new interactions.

In this context, muonic atoms provide an exceptionally sensitive laboratory
for testing fundamental interactions at the picometer scale \citep{proton}.
The muon's large mass, approximately $200$ times that of an electron,
significantly reduces the Bohr radius for muonic atoms. This reduction means
that the orbiting muon is much closer to the nucleus. As a result, there is an
enhanced sensitivity to effects related to nuclear size, quantum
electrodynamics (QED) corrections, and any potential new short-range
interactions. This increased sensitivity was famously demonstrated by muonic
hydrogen spectroscopy, which revealed a significant discrepancy in the
proton's charge radius relative to values obtained from electronic hydrogen
and electron-proton scattering---a persistent tension known as the
\textquotedblleft proton radius puzzle\textquotedblright%
\ \citep{nature,science}. More recently, a high-precision measurement of the
Lamb shift in the muonic helium-$4$ ion $\left(  \mu^{4}\text{He}\right)
^{+}$ yielded a new, highly accurate value for the alpha particle's charge
radius \citep{Krauth2021}. While this result is consistent with
electron-scattering data \citep{Sick2008}, the underlying discrepancy with the
proton radius puzzle provides a compelling motivation to use this atomic
system to search for physics beyond the SM \citep{Dahia2024}.

In this work, we utilize the remarkable precision of muonic atom spectroscopy
to place new constraints on non-Newtonian gravitational corrections using
complementary approaches. We begin by using the precise measurement of the
$2S-2P$ Lamb shift in the muonic helium-4 ion to constrain the
spin-independent Yukawa-type interaction. Further, we leverage the precise
determination of the deuteron-proton squared charge radii difference
($r_{d}^{2}-r_{p}^{2}$), obtained from muonic hydrogen and deuterium
spectroscopy, to establish independent bounds. Although individually the radii
$r_{p}$ and $r_{d}$ measured from muonic atoms deviate from the values
obtained from conventional ones, the isotope shifts inferred from electronic
and muonic spectroscopy coincide with high precision. This agreement enables
the determination of new constraints on the anomalous behavior of gravity at
the short-distance scale. Finally, motivated by the MTV-G experiment, we extend
our analysis to the Post-Newtonian formalism, using the muonic helium
fine-structure splitting data to constrain spin-dependent gravitational effects.

The structure of this work is organized as follows: Sec. II details the
derivation of constraints derived from the muonic helium-$4$ Lamb shift, while
Sec. III presents the complementary analysis based on the muonic
hydrogen-deuterium isotope shift. In Sec. IV, we extend our investigation to
the Post-Newtonian formalism to constrain spin-dependent gravitational effects
by studying the influence of the gravitational spin-orbit coupling on the
$2P$-level splitting of muonic helium. We present our concluding remarks in
Sec. V.

\section{Muonic helium-4 ion bounds}

The analysis of highly accurate atomic transitions provides a powerful method
for probing modifications to the gravitational interaction at the Angstrom
scale. In this context, precise measurements of the $\text{2P--2S}$ transition
in a muonic atom have been used to search for traces of physics beyond the
Standard Model (BSM) \citep{Dahia2024}. The spectroscopy of muonic ions also
allows straightforward investigations of nuclear structure. For instance,
measurements of the $\text{2P--2S}$ transitions in the muonic helium-$4$ ion
$\left(  \mu^{4}\text{He}\right)  ^{+}$ were primarily carried out to
determine the root-mean-square (r.m.s.) charge radius of the alpha particle,
$r_{\alpha}$ \citep{Krauth2021}. As the new value is consistent with the
charge radius obtained from electron-helium scattering experiments
\citep{Sick2008}, it can be utilized to establish independent limits for
non-standard theories.

The theoretical energy splitting for the $\text{2P$_{1/2}$\textendash2S}$
transition of the $\left(  \mu^{4}\text{He}\right)  ^{+}$ion is given by
\citep{Krauth2021}:
\begin{align}
\Delta E_{\text{2P$_{1/2}$}\text{\textendash}\text{2S}}^{\text{theo}}=  &
\left[  1677.690\left(  292\right)  -106.220\left(  8\right)  \times\left(
\frac{r_{\alpha}}{\text{fm}}\right)  ^{2}\right]  \text{meV}. \label{(He4)1}%
\end{align}
It is noteworthy that only the second term explicitly includes the finite-size
effect, as it is proportional to the squared r.m.s. charge radius of the
nucleus. Furthermore, the total theoretical uncertainty to this transition is
$\delta E_{\text{theo}}=0.292\text{ meV}$. On the other hand, the
experimentally measured Lamb shift is $\Delta E_{\text{2P$_{1/2}$%
}\text{\textendash}\text{2S}}^{\text{exp}}=1378.521\left(  48\right)
\text{meV}$ \citep{Krauth2021}. Comparing theory and experiment yields a
combined uncertainty $\delta E=\sqrt{\delta E_{\text{theo}}^{2}+\delta
E_{\text{exp}}^{2}}=0.296\text{ meV}$. Any potential deviation in the atomic
energy levels arising from BSM physics must be smaller than this total uncertainty.

To estimate the non-Newtonian contribution to the energy levels, we consider
the gravitational interaction between the muon and the helium-$4$ nucleus,
described by the perturbation $H_{G}=-GM_{\alpha}m_{\mu}/r\left(
1-e^{-r/\lambda}\right)  $ to the atomic Hamiltonian. Thus, to leading order,
the gravitational shift on the energy difference between the $2\text{P}_{1/2}$
and $2\text{S}$ levels is:
\begin{equation}
E_{G}=\left\langle H_{G}\right\rangle _{\text{2P}}-\left\langle H_{G}%
\right\rangle _{\text{2S}}=\alpha\frac{Gm_{N}m_{\mu}a_{0}\left(  \mu
^{4}\text{He}^{+}\right)  \lambda^{2}}{2\left[  a_{0}\left(  \mu^{4}%
\text{He}^{+}\right)  +\lambda\right]  ^{4}},
\end{equation}
where $a_{0}(\mu^{4}\text{He}^{+})$ is the Bohr radius of the muonic helium
ion. The Bohr radius for muonic helium, $a_{0}(\mu^{4}\text{He}^{+}%
)\simeq10^{-13}\text{ m}$, is about $200$ times smaller than its electronic
counterpart, making the system highly sensitive to new interactions at this
short-distance scale. We now aim to impose stringent constraints on the
parameter space $\left(  \lambda,\alpha\right)  $ by requiring that this new
physics contribution remains undetected within the combined experimental and
theoretical precision, i.e., $|E_{G}|<\delta E$. As we will see, the resulting
exclusion limits on the Yukawa parameters $\left(  \lambda,\alpha\right)  $
derived from this single-ion analysis are more restrictive than those bounds
established by electronic spectroscopy below the picometer scale. However, for
a comprehensive view, these constraints will be presented and discussed in the
next section alongside those derived from the isotope shifts in muonic
hydrogen and deuterium, enabling a direct comparison of several bounds.

\section{Deuteron-Proton Squared Charge Radii Difference Constraints}

Complementing the single-ion approach from the previous section, further
independent constraints on new short-range interactions can be derived from
precise spectroscopic measurements of the isotope shift between muonic
hydrogen $\left(  \mu H\right)  $ and muonic deuterium $\left(  \mu D\right)
$ \citep{Dahia2024}. In this context, the difference of the squared charge
radii of the deuteron $\left(  r_{d}\right)  $ and the proton $\left(
r_{p}\right)  $ can be extracted with remarkable precision from the Lamb shift
in muonic atoms. This difference measurement, $r_{d}^{2}-r_{p}^{2}$, is
particularly prominent as many theoretical uncertainties, such as those
related to nuclear structure and QED corrections, partially cancel, providing
a powerful test of the Standard Model.

The presence of a new Yukawa-type interaction, as described in the previous
section, would contribute an additional energy shift to the Lamb shift $E_{L}$
of both muonic atoms \citep{Pachucki:2022tgl}
\begin{equation}
E_{\text{L}}\left(  \text{theo}\right)  =E_{\text{QED}}+\mathcal{C}r_{N}%
^{2}+E_{\text{NS}}+E_{\text{G}},
\end{equation}
where $E_{\text{QED}}$ encompasses the QED contributions for a point-like
nucleus, the second term is the finite-size nuclear contribution,
$E_{\text{NS}}$ accounts for nuclear structure effects, and $E_{\text{G}}$ is
the additional shift induced by the hypothetical non-Newtonian gravitational
correction. The inferred mean square charge radius, $r_{N}^{2}$, for a given
nucleus $N$ (where $N=p$ or $d$) is determined from a direct comparison
between $E_{L}\left(  \text{theo}\right)  $ and $E_{L}\left(  \text{exp}%
\right)  $. Consequently, in this context, the difference of squared nuclear
charge radii obtained from the weighted isotope shift in muonic atoms
$E_{\text{L}}|_{\text{$\mu$D}}/\mathcal{C_{\text{$\mu$D}}}-E_{\text{L}%
}|_{\text{$\mu$H}}/\mathcal{C_{\text{$\mu$H}}}$ may be written as
\begin{equation}
r_{d}^{2}-r_{p}^{2}=\left(  r_{d}^{2}-r_{p}^{2}\right)  |_{\text{SM}}-\left(
\frac{E_{\text{G}}\left(  \mu D\right)  }{\mathcal{C}_{\text{$\mu$D}}}%
-\frac{E_{\text{G}}\left(  \mu H\right)  }{\mathcal{C}_{\text{$\mu$H}}%
}\right)  , \label{diff}%
\end{equation}
where $\left(  r_{d}^{2}-r_{p}^{2}\right)  |_{\text{SM}}=3.8200\left(
7\right)  \left(  30\right)  \text{fm}^{2}$ is the SM prediction for this
difference, and $\mathcal{C}_{\mu\text{D}}=-6.1074\ \text{meV/fm}^{2}$ and
$\mathcal{C}_{\mu\text{H}}=-5.2259\ \text{meV/fm}^{2}$ are known coefficients
\citep{Pachucki:2022tgl}. By considering the gravitational interaction
parametrized only by the Yukawa term, one gets $E_{G}=\left\langle
H_{G}\right\rangle _{\text{2P}}-\left\langle H_{G}\right\rangle _{\text{2S}}$,
with $H_{G}=-Gm_{N}m_{\mu}\alpha e^{-r/\lambda}/r$. We can establish
constraints on the parameter space $\left(  \lambda,\alpha\right)  $ by
requiring that the theoretical prediction from equation (\ref{diff}) must be
compatible with the combined experimental and theoretical uncertainties
$\delta r_{\text{(exp+theo)}}^{2}=\sqrt{0.0007^{2}+0.0030^{2}}\simeq
0.0031\text{fm}^{2}$:
\begin{equation}
\left\vert \frac{E_{\text{G}}\left(  \mu D\right)  }{\mathcal{C}_{\text{$\mu
$D}}}-\frac{E_{\text{G}}\left(  \mu H\right)  }{\mathcal{C}_{\text{$\mu$H}}%
}\right\vert <\delta r_{\text{(exp+theo)}}^{2}.
\end{equation}
Furthermore, an explicit calculation gives
\begin{equation}
\alpha<\frac{2\delta r_{\text{(exp+theo)}}^{2}}{Gm_{\mu}}\left\vert
-\dfrac{m_{d}a_{0}\left(  \mu\text{D}\right)  }{\mathcal{C}_{\mu D}}\left(
\dfrac{\lambda^{2}}{\left(  \lambda+a_{0}\left(  \mu\text{D}\right)  \right)
^{4}}\right)  +\dfrac{m_{p}a_{0}\left(  \mu\text{H}\right)  }{\mathcal{C}%
_{\mu\text{H}}}\left(  \dfrac{\lambda^{2}}{\left(  \lambda+a_{0}\left(
\mu\text{H}\right)  \right)  ^{4}}\right)  \right\vert ^{-1}, \label{bound2}%
\end{equation}
where $a_{0}\left(  \mu\text{D}\right)  $ and $a_{0}\left(  \mu\text{H}%
\right)  $ represent the Bohr radii of muonic deuterium and muonic hydrogen, respectively.

Using the CODATA 2018 \citep{Tiesinga:2021myr} recommended values for all
quantities in Eq. (\ref{bound2}), we obtain the corresponding exclusion region
in the Yukawa parameter space. The results are presented in Figure 1. For
comparison, the figure also shows the constraint derived from the muonic
helium-$4$ ion analysis in Section II (i.e., $\left\vert E_{G}\right\vert
<\delta E$). This combined analysis shows that for Yukawa ranges
$\lambda\lesssim10^{-12}\text{ m}$, the constraints derived from muonic atom
spectroscopy are significantly more stringent than those from electronic
hydrogen measurements. Notably, the isotope shift method (blue dashed curve)
provides the most restrictive limits in the region such that $\lambda
\lesssim10^{-13}\text{ m}$.

\begin{figure}[t]
\centering{}\includegraphics[scale=0.6]{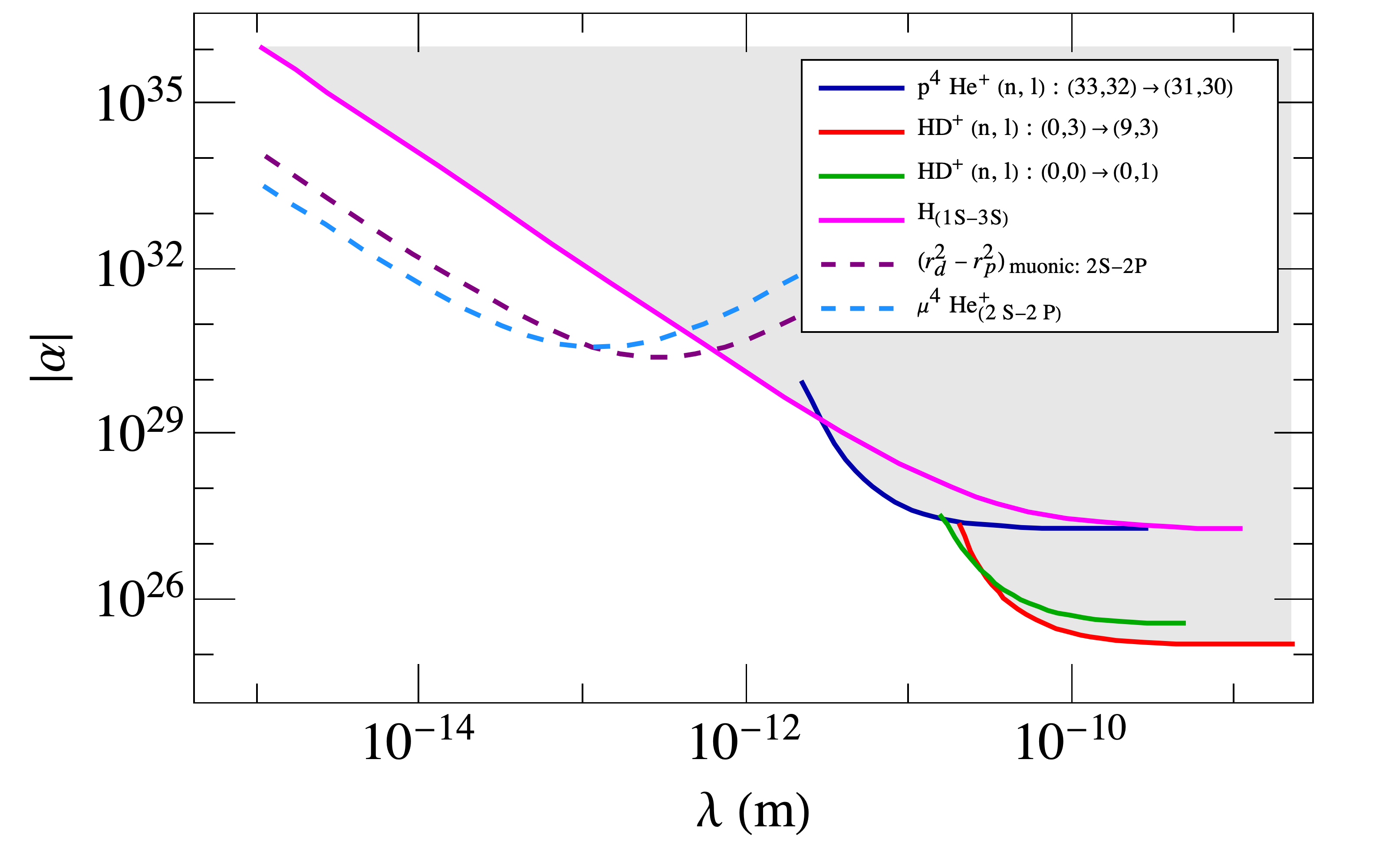}
\caption{\label{fig1}Exclusion limits for the spin-independent Yukawa strength parameter
$\alpha$ as a function of the interaction range $\lambda$. The new constraints
derived in this work, obtained from the $\mu^{4}\text{He}^{+}$ $2S-2P$ Lamb
shift (Sec. II) and the $r_{d}^{2}-r_{p}^{2}$ isotope shift (Sec. III), are
shown as dashed lines. These are compared with the existing $90\%$ C.L.
spectroscopic bounds from atomic transitions (such as H(1S-3S) and p$^{4}%
$He$^{+}$) \citep{lemos4} and molecular HD$^{+}$ transitions
\citep{Germann:2021koc}. The shaded area denotes the excluded region.}%
\end{figure}

The enhanced bounds of the isotope-shift method in this context arise from the
partial cancellation of systematic uncertainties and nuclear structure effects
that influence both hydrogen and deuterium. This cancellation yields a more
precise test of new physics. The results presented in Figure \ref{fig1}
clearly show an improvement provided by muonic spectroscopy in constraining
the anomalous behavior of the gravitational interaction on the subatomic scale.
These analyses, as we have emphasized, were based on a spin-independent
interaction affecting the Lamb shift. Nevertheless, the available
spectroscopic data of muonic helium also contains information about the
$2P_{1/2}-2P_{3/2}$ transition, which is more sensible to spin-dependent
phenomena. Motivated by this characteristic, in the next section, we will
explore this transition to investigate possible modifications of the
gravitational spin-orbit coupling from the analysis of the fine-structure
splitting in muonic atoms.

\section{Post-Newtonian Formalism and Gravitational Spin-Orbit Coupling}

The preceding sections established new constraints on a speculative,
spin-independent Yukawa interaction by analyzing the Lamb shift in muonic
atoms. We now broaden our investigation to consider a more general class of
gravitational modifications, often described within the Post-Newtonian (PPN)
framework \citep{Will:2014kxa}. This formalism allows for the temporal
($\varphi$) and spatial ($\tilde{\varphi}$) components of the gravitational
potential to differ, giving rise to new physical phenomena not accessible
through the Lamb shift alone \citep{lemos4}.

Specifically, a difference between these potentials generates an anomalous
gravitational spin-orbit (GSO) interaction for a particle moving in the
modified spacetime. This effect is negligible in the Lamb shift, which is
dominated by scalar contributions\footnote{ The temporal component of the
metric behaves as a scalar quantity with respect to purely spatial coordinates
transformation.}, but it directly impacts energy levels that differ only by
the orientation of the particle's spin. The fine-structure splitting between
the $2P_{3/2}$ and $2P_{1/2}$ states is therefore a suitable system to search
for this spin-dependent physics. In this section, we derive a new and
independent set of constraints, complementary to the scalar bounds from
Sections II and III, by analyzing the GSO effect on the fine-structure of
muonic helium.

To investigate these effects at the atomic scale, we consider the
gravitational field produced by a nucleus as static and spherically symmetric.
In this context, one can write the metric of the spacetime as \citep{lemos4}:
\begin{equation}
ds^{2}=-c^{2}\left(  1+\frac{2\varphi}{c^{2}}\right)  dt^{2}+\left(
1-\frac{2\tilde{\varphi}}{c^{2}}\right)  (dx^{2}+dy^{2}+dz^{2}),
\label{eq:metrica_ds}%
\end{equation}
where $\varphi(r)$ and $\tilde{\varphi}(r)$ represent the gravitational
potentials. The potential $\tilde{\varphi}$ is directly associated with the
curvature of the spatial section of spacetime, and any deviation from the
potential $\varphi$ is parametrized by $\gamma$ in the PPN formalism. Although
General Relativity suggests that $\varphi=\tilde{\varphi}$, we aim to explore
the possibility that they differ at short length scales, which would be
interpreted as a signal of new physics.

We model this eventual deviation with a short-range Yukawa-type interaction.
Thus, the gravitational potentials are given by:
\begin{align}
\varphi(r)  &  =-\frac{GM}{r}(1+\alpha e^{-r/\lambda}),\\
\tilde{\varphi}(r)  &  =-\frac{GM}{r}(1+\tilde{\alpha}e^{-r/\lambda}),
\end{align}
where $\alpha$ and $\tilde{\alpha}$ are dimensionless parameters that set the
strength of the new interaction responsible for the correction.

The post-Newtonian potential $\tilde{\varphi}$ does not contribute to the
atomic Lamb shift with the same order of magnitude as the potential $\varphi$.
However, in the gravitational spin-orbit (GSO) coupling, these two potentials
play equivalent roles. Here, we intend to examine the effect of a non-standard
spatial potential $\tilde{\varphi}$ at short length by analyzing the
fine-structure splitting between the $2P_{3/2}$ and $2P_{1/2}$ levels of
muonic helium, which is particularly sensitive to such spin-orbit coupling.
The GSO coupling will induce an anomalous energy shift $\Delta E_{\text{GSO}}$
between these two states. The magnitude of this shift can be calculated by
finding the difference between the expectation values of the GSO Hamiltonian,
$H_{\text{GSO}}$, for each state:
\begin{equation}
\Delta E_{\text{GSO}}=\left\langle H_{\text{GSO}}\right\rangle _{2P_{3/2}%
}-\left\langle H_{\text{GSO}}\right\rangle _{2P_{1/2}}.\label{eq:delta_Egso}%
\end{equation}
Using the Foldy-Wouthuysen formalism, we can find the operator that describes
this interaction. For the spherically symmetric potentials under
consideration, the Hamiltonian can be expressed in the standard form of
spin-orbit coupling \citep{Silenko:2004ad,Fischbach:1981ne}:
\begin{equation}
H_{\text{GSO}}=\frac{1}{mc^{2}}\left[  \frac{1}{r}\frac{d}{dr}\left(
\frac{\varphi}{2}+\tilde{\varphi}\right)  \right]  (\vec{S}\cdot\vec
{L}).\label{eq:hamiltonian_gso}%
\end{equation}
Note that if we only consider the standard Newtonian potential ($\alpha
=\tilde{\alpha}=0$), the term in the brackets is proportional to $1/r^{3}$,
and we recover the standard GSO coupling of GR. Thus, the new physics is
contained entirely within the Yukawa terms.

The calculation of the expectation value of $H_{\text{GSO}}$ requires
evaluating two separate terms: the radial integral and the spin-angular part.
For the radial component, we use the $2P$ state wave-function ($n=2,l=1 $) for
the muonic helium ion:
\begin{equation}
R_{21}(r)=\frac{1}{\sqrt{24a_{0}^{3}}}\frac{r}{a_{0}}e^{-r/2a_{0}},
\label{eq:radial_21}%
\end{equation}
where $a_{0}$ is the appropriate Bohr radius. The derivative of the potentials
containing only the hypothetical new physics is:
\begin{equation}
\frac{d}{dr}\left(  \frac{\varphi}{2}+\tilde{\varphi}\right)  =GM\left(
\frac{\alpha}{2}+\tilde{\alpha}\right)  e^{-r/\lambda}\left(  \frac{1}{r^{2}%
}+\frac{1}{r\lambda}\right)  .
\end{equation}

For the spin-angular part, we use the identity $\vec{S}\cdot\vec{L}=\frac
{1}{2}(J^{2}-L^{2}-S^{2})$, so that the expectation values for the $2P_{3/2}$
($j=3/2$) and $2P_{1/2}$ ($j=1/2$) states are $\frac{1}{2}\hbar^{2}$ and
$-\hbar^{2}$, respectively. The difference between them is therefore:
\begin{equation}
\left<  \vec{S}\cdot\vec{L}\right>  _{j=3/2}-\left<  \vec{S}\cdot\vec
{L}\right>  _{j=1/2}=\frac{3}{2}\hbar^{2}.
\end{equation}
Finally, by combining the results of the radial and angular parts, we can find
the expression for the anomalous energy shift:
\begin{equation}
\Delta E_{SO}=\dfrac{GM_{\alpha}}{m_{\mu}c^{2}}\left(  \dfrac{\alpha}%
{2}+\Tilde{\alpha}\right)  \left[  \dfrac{1}{24a_{0}^{3}}\dfrac{\lambda
^{2}\left(  3a_{0}+\lambda\right)  }{\left(  a_{0}+\lambda\right)  ^{2}%
}\right]  \dfrac{3}{2}\hbar^{2}.
\end{equation}

To establish the constraints, we impose the condition that this new
contribution cannot exceed the total uncertainty of the transition energy,
$|\Delta E_{\text{GSO}}|<\delta E_{\text{total}}$. The reported experimental
and theoretical uncertainties are, respectively,
\begin{align}
\delta E_{\text{exp}}  &  =\pm0.096\text{ meV},\\
\delta E_{\text{the}}  &  =\pm0.0003\text{ meV}.
\end{align}

\begin{figure}[t]
\centering\includegraphics[scale=0.6]{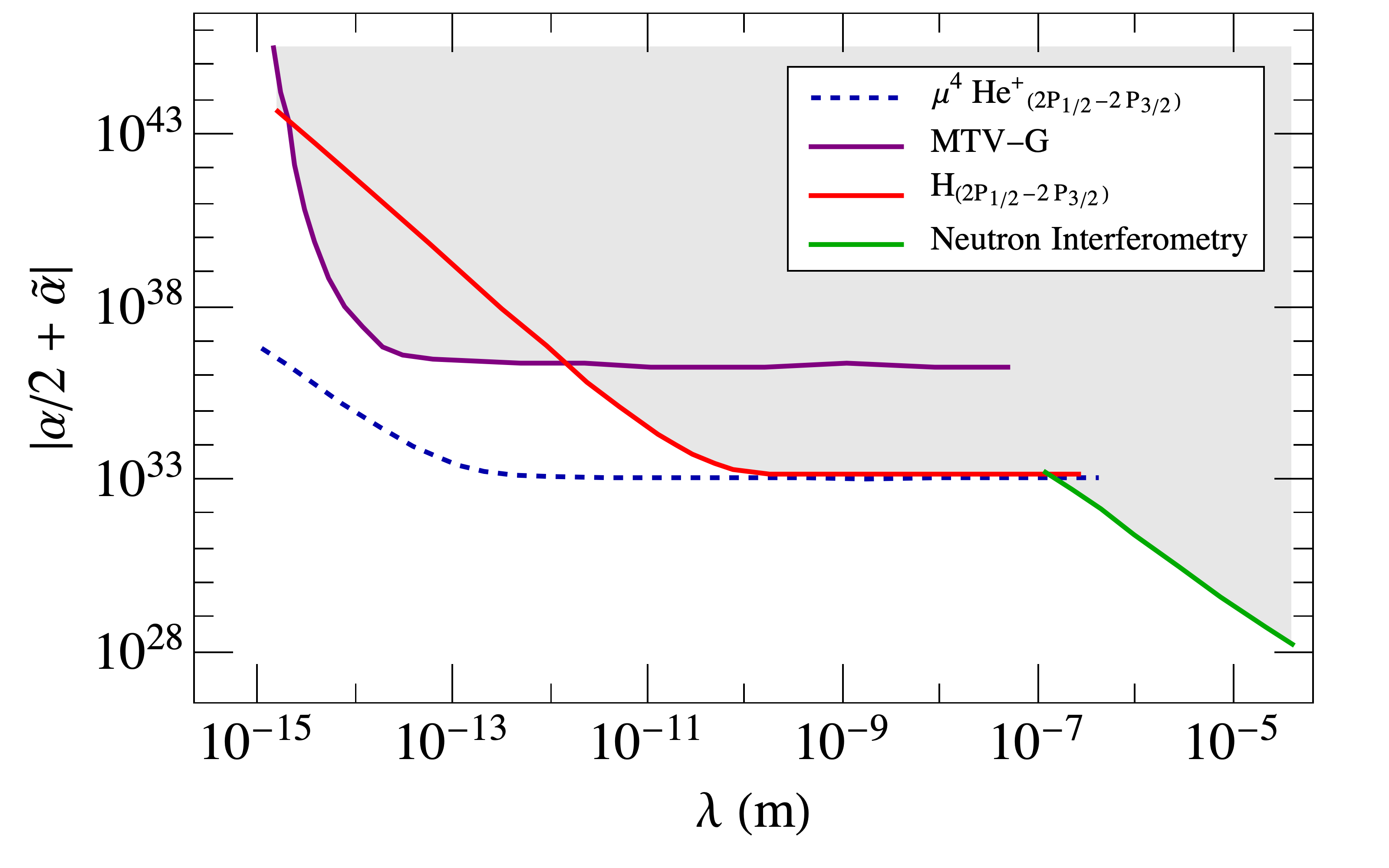}%
\caption{\label{fig2}Constraints on the PPN parameter combination $(\alpha/2+\tilde
{\alpha})$ at $68\%$ C.L.. The new bound derived in this work from the muonic
helium-4 fine-structure splitting is depicted by the dashed blue line. This
result is compared with existing limits from neutron interferometry
\citep{Rocha:2021zgw} (green line), MTV-G experiment data
\citep{murata,Tanaka:2014jfa,Tanaka:2013ika} (purple line), and from the
analysis of the fine-structure of the hydrogen $2P$-level \citep{lemos4} (red
line). The parameter space within the shaded region is excluded. }
\end{figure}

Considering these errors to be independent, the total uncertainty is $\delta
E_{\text{total}}=\sqrt{\delta E_{\text{exp}}^{2}+\delta E_{\text{the}}^{2}%
}\approx0.096\text{ meV}$. This condition sets the allowed limits on the
parameter space, as illustrated in Fig. \ref{fig2}.

From Fig. \ref{fig2}, we observe that the constraints obtained from the muonic
helium-4 ion on the parameter combination $(\alpha/2+\tilde{\alpha})$ are
significantly stronger than those from electronic hydrogen and the MTV-G
experiment for Yukawa ranges $\lambda<10^{-10}\text{ m}$. This result once
again highlights the remarkable prospect of muonic atoms as a laboratory for
testing fundamental physics and searching for deviations from General
Relativity at the microscopic scale.

\section{Concluding Remarks}

High-precision spectroscopy of muonic atoms provides a powerful avenue for
searching for new physics at short distances (below a nanometer). In this
work, we have leveraged the remarkable sensitivity of these systems to derive
improved constraints on hypothetical non-Newtonian gravitational interactions,
often parametrized by a Yukawa-type potential. We conducted a comprehensive
investigation using three distinct, complementary spectroscopic probes to
cover both spin-independent and spin-dependent effects.

First, we explored constraints on spin-independent interactions, characterized
by the strength parameter $\alpha$. By analyzing the precisely measured
$2S-2P$ Lamb shift in the muonic helium-4 ion (Sec. II) and the
deuteron-proton squared charge radii difference from the muonic
hydrogen-deuterium isotope shift (Sec. III), we established new exclusion
regions in the $\left(  \alpha,\lambda\right)  $ parameter space. Our results
show that the isotope shift method provides the most restrictive bounds for
very short interaction ranges $\left(  \lambda\lesssim10^{-13}\text{
m}\right)  $, owing to the partial cancellation of theoretical uncertainties.
For a broad range of $\lambda$ below the picometer scale, these constraints
significantly surpass those derived from electronic atom spectroscopy.

Furthermore, we extended our investigation to explore spin-dependent
gravitational effects within the Post-Newtonian formalism. By analyzing the
effect of the gravitational spin-orbit (GSO) coupling on the $2P_{3/2}%
-2P_{1/2}$ fine-structure splitting in muonic helium (Sec. IV), we impose
limits on the PPN parameter combination $(\alpha/2+\tilde{\alpha})$. We have
shown that for interaction ranges $\lambda<10^{-10}\text{ m}$, these bounds
are more stringent than those from other experimental techniques, including
neutron interferometry and MTV-G searches, demonstrating the competitive
advantage of this approach.

\begin{acknowledgments}
We would like to thank CNPq, CAPES, and FAPESQ-PB, for partial financial
support. J. E. J. M. acknowledges support from FAPESQ-PB (Grant No. 16--2/2022).
\end{acknowledgments}

\end{document}